\documentclass[superscriptaddress,showpacs,preprintnumbers,nofootinbib,twocolumn]{revtex4}

\usepackage{amsmath,amssymb,graphicx}

\begin{document}

\newcommand*{\pku}{School of Physics and State Key Laboratory
of Nuclear Physics and Technology, Peking University, Beijing
100871, China}\affiliation{\pku}

\newcommand*{\CHEP}{Center for High Energy
Physics, Peking University, Beijing 100871,
China}\affiliation{\CHEP}

\title{Lorentz violation induced vacuum birefringence and its astrophysical consequences\footnote{This is the short version
for publication in PRD. For a more detailed long version, please see
Version 1 of this arXiv paper arXiv:1104.4438v1}}

\author{Lijing Shao}\affiliation{\pku}
\author{Bo-Qiang Ma}\email[Corresponding author.
Electronic
address:~]{mabq@pku.edu.cn}\affiliation{\pku}\affiliation{\CHEP}

\begin{abstract}
In the electromagnetism of loop quantum gravity, two helicities of a
photon have different phase velocities and group velocities, termed
as ``vacuum birefringence''. Two novel phenomenons, ``peak
doubling'' and ``de-polarization'', are expected to appear for a
linearly polarized light from astrophysical sources. We show that
the criteria to observe these two phenomenons are the same. Further,
from recently observed $\gamma$-ray polarization from Cygnus X-1, we
obtain an upper limit $\sim 8.7\times10^{-12}$ for Lorentz-violating
parameter $\chi$, which is the most firm constraint from well-known
systems. We also suggest to analyze possible existence of ``peak
doubling'' through Fermi LAT GRBs.
\end{abstract}

\pacs{11.30.Cp, 04.60.-m, 78.20.Fm, 95.85.Pw}

\maketitle


Symmetries are important ingredients in modern physics, and among
which Lorentz symmetry is preeminently fundamental and profound.
However, in searches of quantum gravity (QG), Lorentz violation (LV)
emerges in many theoretical frameworks that try to conciliate
apparent conflictions between standard model and general
relativity~\cite{a98nat,ck98prd,gp99prd,a02ijmpd}. As it offers a
valuably observational window on QG, LV has stimulated lots of
experimental works. Hitherto, LV parameters have been severely
constrained from astrophysical observations and terrestrial
experiments on various
species~\cite{m05lrr,jlm06ap,kr08,lm09arnps,sm10mpla,sxm10app,xm09prd}.
However, no firm Lorentz-violating phenomenon has been confirmed
yet.

Amongst phenomenological works, vacuum birefringence (VB), which
shows great sensitivity to LV physics, is extensively
studied~\cite{ck98prd,gp99prd,cfj90prd,gk01prd,mp03prl,jlm04prl,f07mnras,m08prd,km08apjl,xsm10epjc,s11}.
VB is an analogy with birefringence within anisotropic medium, where
left-handed and right-handed modes of light travel with different
phase velocities and group velocities. It can arise from many
parity-violating theories, including Chern-Simons
terms~\cite{cfj90prd,afg10apj}, loop quantum
gravity~\cite{gp99prd,gk01prd}, Lorentz-violating effective field
theories~\cite{ck98prd,mp03prl,jlm04prl,km08apjl,xsm10epjc}.
Lorentz-violating effects can modify phase velocities and group
velocities of two oppositely circularly polarized modes, and they
individually get a modification with an opposite sign. The
modification is believed to be suppressed by some powers of the
Planck length $l_{\rm Pl} \equiv \sqrt{G\hbar/c^3} \simeq 1.6 \times
10^{-35}$~m.\footnote{However, there are also arguments that a new
fundamental scale might appear rather than the conventional Planck
scale, see e.g., Ref.~\cite{sm10} and references therein.}

As a consequence of VB, an originally linearly polarized light,
which composes of left-handed and right-handed modes, from
astrophysical and cosmological distance, will manifest ``peak
doubling'' or ``de-polarization'' features when entering the
observer~\cite{gp99prd,gk01prd,jlm04prl,f07mnras,m08prd,s11}. From
this scenario, the Lorentz-violating parameter is constrained to a
great precision from observations of polarized lights from the Crab
Nebula~\cite{m08prd} and $\gamma$-ray bursts
(GRBs)~\cite{jlm04prl,f07mnras,s11}. In addition to probe LV
physics, VB can also serve to distinguish parity-violating theories
from those of even parity, like foamy spacetime~\cite{a98nat} and
doubly special relativity~\cite{a02ijmpd,zsm11app}.

In this report, we utilize the Lorentz-violating electromagnetism in
loop quantum gravity~\cite{gp99prd}. By adopting an Ansatz
accounting for differences in both phase velocities and group
velocities, we obtain propagation behaviors and Stocks parameters of
a linearly polarized light from cosmological distance. We show that
the criteria to observe ``peak doubling'' and ``de-polarization''
are the same. By utilizing our derived formula to recently observed
polarization of $\gamma$-rays from Cygnus X-1, we obtain an upper
limit $\sim 8.7\times10^{-12}$ for Lorentz-violating parameter
$\chi$, which turns out to be the most firm constraint from
well-known systems, though a little looser than that from the
distance-estimated GRB 041219A~\cite{s11}. Further, we re-propose
the idea to analyze possible existence of peak doubling in light
curves of most energetic Fermi LAT GRBs. In the paper, the
convention $\hbar=c=1$ is used.


In the picture of semi-classical spacetime with ``polymer-like''
structure that emerges in the loop quantum gravity, Gambini and
Pullin derived the modified Maxwell equations~\cite{gp99prd},
$\partial_t \vec{E}  =   \nabla \times \vec{B} + 2 \chi l_{\rm Pl}
\nabla^2 \vec{B}$ and $\partial_t \vec{B} = - \nabla \times \vec{E}
- 2 \chi l_{\rm Pl} \nabla^2 \vec{E}$, which break Lorentz boost
symmetry as well as parity, while preserving rotation symmetry. From
modified Maxwell equations, it is straightforward to get the
modified dispersion relation for photons, $\Omega_\pm = |\vec{k}|
\mp 2 \chi l_{\rm Pl} |\vec{k}|^2$, where $\Omega_\pm$ are
frequencies for left-handed and right-handed modes. A similar
dispersion relation, $ \Omega_\pm = |\vec{k}| \mp \xi l_{\rm Pl}
|\vec{k}|^2 $, can be attained from an effective field theory with
Lorentz-violating dimension-5 operators for the photon
sector~\cite{mp03prl}.

Now from the modified dispersion relation, the phase velocity
$v^{\rm p}$ and group velocity $v^{\rm g}$ of photons become
\begin{equation}
v^{\rm p}_\pm \equiv \frac{\Omega_\pm}{|\vec{k}|} = 1 \mp 2\chi
l_{\rm Pl} |\vec{k}|, ~ v^{\rm g}_\pm \equiv
\frac{\partial\Omega_\pm}{\partial|\vec{k}|} = 1 \mp 4\chi l_{\rm
Pl}|\vec{k}|,
\end{equation}
respectively, and noticeably they are both helicity dependent,
namely vacuum birefringent.


We consider a fully linearly polarized light from astrophysical
sources, whose electrical field $\vec{E}$ is a superposition of two
monochromatic waves with opposite circular polarizations, i.e.,
$\vec{E} = \vec{E}_+(k_+) + \vec{E}_-(k_-)$. The radiation can be
produced from various mechanisms, e.g., through synchrotron
radiation in a region penetrated with well ordered magnetic fields.
This can be achieved in the vicinity of a neutron star, of an active
galactic nucleus (AGN), and of a GRB. For a photon traveling along
$z$-axis with its central frequency $\Omega_0$, the wavenumbers for
two modes are $k_\pm = \Omega_0 (1 \pm 2 \chi l_{\rm Pl}\Omega_0 )$.
Assuming a Gaussian wave packet with a width $\Delta$ in space, we
have~\cite{gk01prd}
\begin{equation}\label{packet}
\vec{E}_\pm \propto {\rm Re} \left\{  \exp{ [i (\Omega_0 t - k_\pm z
)]} \exp{\left[ -\frac{(z-v^{\rm g}_\pm t)^2 }{\Delta^2} \right]}
\hat{\rm e}_\pm \right\},
\end{equation}
where $\hat{\rm e}_\pm \equiv \hat{\rm e}_1 \pm i\hat{\rm e}_2$.

Conventionally, two modes arrive at the earth at the same time
$t_0=z$, hence we are able to detect a superposition, i.e., a
linearly polarized light. However, with LV effects, VB is induced,
and their times of arrival can be different. To focus on the
detection epoch, let us notate $z=t+\delta t$. Now bigger $\delta t$
means earlier arrival.

\begin{figure}
\includegraphics[width=9cm]{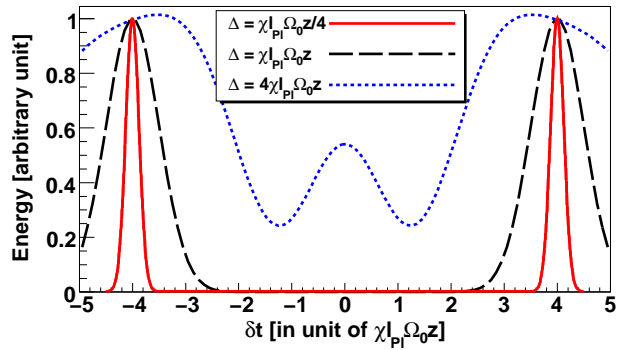}\caption{The
radiation arrives at the earth versus the ``arrival time'' $\delta
t$.}\label{energy}
\end{figure}

Ref.~\cite{gp99prd} noticed that these two modes will be separated
by a distance $\sim 8\chi l_{\rm Pl} \Omega_0 z$, hence they arrive
at the earth in sequence. The energy, $\propto |\vec{E}|^2$, arrives
at the earth versus the ``arrival time'' $\delta t$ is illustrated
in Fig.~\ref{energy} for three different widths in space, $\Delta =
\chi l_{\rm Pl} \Omega_0 z/4$, $\chi l_{\rm Pl} \Omega_0 z$, and
$4\chi l_{\rm Pl} \Omega_0 z$. We adopt $\chi l_{\rm Pl} \Omega_0^2
z = 1$ in the calculation. Typically, we have $\chi l_{\rm Pl}
\Omega_0^2 z \simeq 4.8
\frac{\chi}{10^{-14}} %
\frac{l_{\rm Pl}}{10^{-28}~{\rm eV}^{-1}} %
(\frac{\Omega_0}{100~{\rm keV}})^2 %
\frac{z}{10^{10}~{\rm l.y.}}$. For comparisons, we also calculate
$\chi l_{\rm Pl} \Omega_0^2 z = 10$ and $\chi l_{\rm Pl} \Omega_0^2
z = 100$ cases.\footnote{See arXiv:1104.4438v1 [astro-ph.HE] for
details.} It is shown that the profile is merely modified. From the
figure, we can see that ``peak doubling'' appears when the width
$\Delta$ is small. With increasing $\Delta$, two peaks tend to merge
into one single signal. Hence to detect such a phenomenon, the width
of packet should be smaller than the doubling separation, i.e.,
$\chi l_{\rm Pl} \Omega_0 z > \Delta$.

We can also easily get canonical Stocks parameters for the
light.$^2$ After averaging over time, the degree of linear
polarization $\varpi_L = \exp \left[ - {32 \chi^2 l_{\rm Pl}^2
\Omega_0^2 z^2}/{\Delta^2} \right]$, which is exponentially
suppressed. Hence the original polarization will be smeared
drastically if $\chi l_{\rm Pl} \Omega_0 z > \Delta$, termed as
``de-polarization''. This criterion turns out to be the same as that
for ``peak doubling''.


Most previous ``de-polarization'' analysis bases on a reasoning
that, the observation of polarization indicates that the rotated
angle caused by VB between photons with low energy $\Omega_{0l}$,
and those with high energy $\Omega_{0h}$, is smaller than $\pi$,
i.e., $|2\chi l_{\rm Pl} (\Omega_{0h}^2 - \Omega_{0l}^2)z| \lesssim
\pi$. This is absolutely plausible when the spectrum is flat,
however, when it deviates from a flat one, essential cautions should
be kept in mind, especially when it is steep. The observed
polarization may be due to dominant low energy photons, because the
contribution from high energy ones can be largely suppressed and
contributes insubstantially. In most realistic cases in
astrophysics, the spectrum turns to be decreasing as a function of
energy, $N(\Omega_0) \propto \Omega_0^{-\Gamma}
\textrm{e}^{-\Omega_0/E_0}$, where $E_0$ denotes a possible cutoff,
and $\Gamma$ varies according to different radiation mechanisms and
electron distributions. Typically, $\Gamma = 1 \sim 4$ for
X/$\gamma$-rays. Therefore, the contribution from high energy
photons is indeed minor, compared to the large population of low
energy ones. Cautions should be kept in mind when the cutoff is
obvious and/or the measured polarization is small, say $\lesssim
5\%$. On the other hand, a more convincing result can be drawn after
taking the population of photons into account and convoluting it
with the Ansatz presented here.


\begin{table}\caption{Some previous observational constraints from astrophysics. Numerical factors between our notation from those of
previous literatures on $|\chi|_{\rm upper}$ are
accounted.}\label{tab}
\begin{tabular}{lllll}
\hline\hline%
{Source} \quad& {$z$} \quad& {Energy} \quad& {$\varpi_L$(\%)} \quad& %
$|\chi|_{\rm upper}$ \\
\hline%
3C 256
 & $\mathcal {Z}$$\simeq$$1.82$ & $3000$-$4000$~{\AA} &  $16.4$$\pm$$2.2$  & $5~10^{-5}$~\cite{gk01prd}\\
GRB 020813
 & $\mathcal {Z}$$\simeq$$1.3$ &
$3500$-$8800$~{\AA} & $1.8$-$2.4$ & $1~10^{-7}$~\cite{f07mnras}\\%
GRB 021004
 & $\mathcal {Z}$$\simeq$$2.3$ & $3500$-$8600$~{\AA} & $\lesssim 2$ & $5~10^{-8}$~\cite{f07mnras}\\%
GRB 021206
 & $\mathcal {Z}$$\sim$$0.1$ & $0.15$-$2.0$~MeV & $80$$\pm$$20$   & $1~10^{-15}$~\cite{jlm04prl}\\
Crab pulsar
 & $\sim $$2$~kpc & $0.1$-$1$~MeV & $46$$\pm$$10$  & $2~10^{-10}$~\cite{m08prd}\\
GRB 041219A
 & $\mathcal {Z}$$\sim$$0.3$ & $100$-$350$~keV & $63^{+31}_{-30}$, $96^{+39}_{-40}$ & $1~10^{-14}$~\cite{s11}\\
\hline%
\end{tabular}
\end{table}

In Table~\ref{tab}, we list six constraints determined previously,
where three utilize optical/ultraviolet lights, while the other use
$\gamma$-rays. Because the rotated angle depends
quadratically on the energy and only linearly on the distance, 
high energy observations have a big advantage. It can be seen
clearly from the table that the highest energy observation, GRB
021206, could place the most stringent constraint~\cite{jlm04prl}.
However, the observation is refuted later~\cite{rf04mnras,w04apj}.
Hence the most stringent constraint comes from GRB 041219A, whose
``pseudo-redshift'' was estimated to be $\mathcal{Z}\sim0.3$ by
Stecker and the Lorentz-violating parameter is determined to be
$\chi \lesssim 10^{-14}$~\cite{s11}. It is significant, though the
estimated distance of GRB 041219A needs further verification. In
addition, polarized observation from the Crab Nebula constrained
firmly on $\chi$ to be $\lesssim 2 \times 10^{-10}$~\cite{m08prd}.

\begin{figure}
\includegraphics[width=9cm]{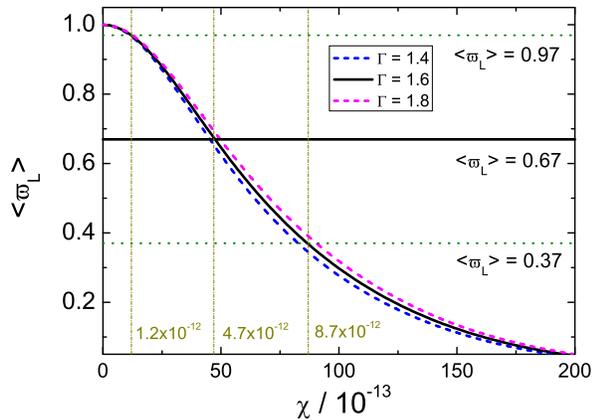}\caption{The spectrally averaged
degree of linear polarization of Cygnus X-1 $\gamma$-ray emissions,
versus the Lorentz-violating parameter $\chi$.}\label{cygnus}
\end{figure}

Recent INTEGRAL/IBIS observation of Cygnus X-1 black hole binary
system found evidently that the $\gamma$-ray emission is largely
polarized with $\langle\varpi_L\rangle = 67 \pm 30\%$ in the energy
band $400$~keV--$2$~MeV~\cite{l11sci}. The spectrally averaged
degree of linear polarization is defined as $\langle \varpi_L
\rangle \equiv \int N(\Omega_0)\varpi_L(\Omega_0) \textrm{d}\Omega_0
/ \int N(\Omega_0) \textrm{d}\Omega_0$, where $N(\Omega_0) \propto
E^{-\Gamma}$ is given by the INTEGRAL observation with a photon
index $\Gamma = 1.6\pm0.2$~\cite{l11sci}. We can roughly estimate
the LV parameter $\chi$ to be around $\sim 10^{-11}$ with a
reasonable width in space $\Delta \sim 1~\textrm{MeV}^{-1}$, and
observational parameter $z = 2.1$~kpc, $\Omega_0 \sim 1$~MeV. To be
more precise, we also perform a detailed analysis taking the
spectral effects into account. The result is shown in
Fig.~\ref{cygnus}, where, as a conservative treatment, $100\%$
polarization at source is assumed. We can see that the photon index
$\Gamma$ has minor effects on the conclusion, so we would use its
central value $\Gamma=1.6$ in the following. Through the observed
value $\langle\varpi_L\rangle = 67 \pm 30\%$, we can infer an upper
limit on $\chi$. The most conservative situation gives
$|\chi|_\textrm{upper}=8.7\times10^{-12}$.

Our upper limit $|\chi|_\textrm{upper}=8.7\times10^{-12}$ is better
than the upper limit from the Crab pulsar, though somehow looser
than that inferred from GRB 041219A. Worthy to mention that, the
difference between Cygnus X-1 and GRB 041219A mainly comes from
distances. However, in the GRB 041219A case, the distance is
estimated through ``pseudo-redshift'', not from direct observations.
In contrast, Cygnus X-1 binary is a much well-known system in
astronomy, hence our upper limit on $\chi$, though seems looser,
actually bases on more firm observations.

Our ``de-polarization'' criterion is somehow different, and only
when $\Delta \simeq 1/\Omega_0$, it turns into $\chi l_{\rm Pl}
\Omega_0^2 z > 1$, then seems similar to that used previously.
However, $\Delta$ is not promised to equal to $1/\Omega_0$, instead,
it should be around $1/\delta \Omega_0$, where $\delta \Omega_0$ is
the uncertainty of the energy of photons, according to Heisenberg's
uncertainty principle. Considering the radiation mechanism of
astrophysical sources, $\delta \Omega_0$ reflects the ``fuzziness''
of the process. The ``fuzziness'' is determined by the environmental
conditions when generating the light, including the irregularities
of magnetic fields, the distribution of electrons, and quantum
mechanical effects. Astrophysicists are putting great efforts to get
these quantities from various observations and inferences. In order
to estimate our criterion, we rewrite it into $\frac{\chi}{10^{-14}} %
\frac{l_{\rm Pl}}{10^{-28}~{\rm eV}^{-1}} %
\frac{\Omega_0}{100~{\rm keV}} %
\frac{\delta\Omega_0}{20~{\rm keV}}%
\frac{z}{10^{10}~{\rm l.y.}} > 1$. We can see that with an organized
environment which provides $\delta\Omega_0 \sim 20~{\rm keV}$, our
criterion is met if we observe a polarized light of $\gtrsim
100~{\rm keV}$ from cosmological distance when $\chi \sim 10^{-14}$.
These conditions are already practical nowadays, therefore, we are
expected to observe or further constrain Lorentz-violating
birefringence from astronomical observations. With upcoming
$\gamma$-ray polarimetric instruments, POET (POlarimeters for
Energetic Transients)~\cite{h08} for an example, we can study VB
better to validate/falsify relevant QG theories.

When VB was considered in the context of loop quantum
gravity~\cite{gp99prd}, BATSE was the best performing satellite in
detecting GRBs. Hence, concerning technical parameters of BATSE,
Gambini and Pullin considered a fictitious GRB at a cosmological
distance, $z \sim 10^{10}$ light years, and $\Omega_0 \sim 200$~keV,
which produces a doubling $\sim 10^{-5}$~s if $\chi$ is of order
$\mathcal{O}(1)$~\cite{gp99prd}. However, this is out of
detectability then. Therefore, main interests of studies of VB were
grasped into possible observations of ``de-polarization'', for it
seems more sensitive to Lorentz-violating parameters.

Nowadays, however, there are several reasons to re-consider studies
of the possible existence of peak doubling. First, as discussed
above, ``de-polarization'' studies can sometimes have problematic
explanations when regarding the non-uniform population of photons
induced by the spectrum. Second, at the time of BATSE, the observed
peak width of GRBs appears to be of order $\sim 0.1~\textrm{s}$,
with features like a rising edge $\sim1$~ms~\cite{gp99prd}. In
contrast, after about ten years, current Fermi LAT instrument has
timing accuracy $< 10 ~\mu\textrm{s}$, and maximum energy
detectability up to $300$~GeV~\cite{lat09apj}. Third, peak doubling
and de-polarization correspond to differences in group velocity and
phase velocity, respectively. There are still debates on theoretical
predictions of these two velocities. Hence, as a different approach
to falsify/verify LV from de-polarization, peak doubling has its own
irreplaceable significance in LV searches. Even if LV is confirmed
from de-polarization observations, it is still largely valuable to
detect peak doubling as a consistent check or an additional study.
In addition, in the high energy band, peak doubling has extra
observational merits, compared with de-polarization. Technically,
timing measurement is somehow easier than polarization measurement
in the high energy band. To pindown polarization properties, we
should have enough events for statistics, in contrast, timing of
detection of high energy photons relays less on statistics. From
another point of view, the doubling signal will be buried in the sea
of photons if the photon flux is too large, and fortunately, at high
energies, we can analyze it more neatly.

Concerning technical progresses made in recent years, we rewrite the
timing separation of peak doubling into $
  8\chi l_{\rm Pl} \Omega_0 z \sim 10^2 \chi
  \frac{l_{\rm Pl}}{10^{-28}~\textrm{eV}^{-1}}
  \frac{\Omega_0}{300~\textrm{GeV}}
  \frac{z}{10^{10}~\textrm{l.y.}}~\textrm{s}$.
With timing accuracy $<10~\mu$s of Fermi LAT, we can have
sensitivity down to $\chi \sim \mathcal{O}(10^{-7})$. This is of the
same order of that from de-polarization constraints from GRB 020813
and GRB 021004~\cite{f07mnras}, though still several orders away
from X/$\gamma$-ray GRB observations (see Table~\ref{tab} for
comparisons). Worthy to mention that, our criterion is automatically
satisfied with Fermi parameters. With eight orders of improvement
over the past ten years, we would be positive to have extra
improvements of magnitudes to detect possible existence of peak
doubling in the future through next generation of satellites, or
through multi-TeV photon observations from ground-based cosmic-ray
observatories.


In summary, Lorentz-violating and parity-violating quantum
gravitational theories predict vacuum birefringence, where two
circularly polarized modes of a linearly polarized light have
different phase velocities and group velocities. Hence, an
originally linearly polarized light produced in astrophysical
processes can manifest new phenomenons when arriving at the observer
after traveling through a cosmological distance. ``Peak doubling''
and ``de-polarization'' are expected to be observed with a
non-vanishing Lorentz-violating parameter of a suitable magnitude.
Inversely, non-observations of these two phenomenons can be used to
constrain the Lorentz-violating parameter.

In this report, we study an Ansatz which concerns both differences
in phase velocities and group velocities of two oppositely
circularly polarized modes in the electromagnetism of loop quantum
gravity. ``Peak doubling'' and ``de-polarization'' phenomenons can
be easily read from our analysis. We also present criteria to
observe ``peak doubling'' and ``de-polarization'' from astrophysical
studies, and these two criteria turn out to be the same. Comparisons
of theoretical criteria and observational practicality are also
presented with some emphasis on possible Fermi LAT observations of
``peak doubling'', which is rarely discussed in literatures. We also
analyze the recently observed polarization of $\gamma$-ray emissions
from Cygnus X-1 black hole, and obtain an upper limit $\chi \lesssim
8.7\times10^{-12}$, which turns out to be the best constraint from
well-known systems. We would expect the quantum-gravitationally
induced vacuum birefringence issue to meet a more distinguishable
situation from upcoming observations of newly planned astronomical
instruments, POET for an example, with abilities to measure
polarizations of high energy photons, or new instruments, Fermi LAT
for an example, with precision to detect rapid variability in the
high energy $\gamma$-ray light curves.

This work is supported by National Natural Science Foundation of
China (11005018, 11021092, 10975003, 11035003).

\end{document}